\documentclass{webofc}
\usepackage[varg]{txfonts}
\usepackage{listings}
\usepackage{hyperref}
\begin{document}

\title{The Impact of Gaia DR1 on Asteroseismic Inferences from Kepler}

\author{\firstname{Travis} \lastname{Metcalfe}\inst{1} \and
        \firstname{Orlagh} \lastname{Creevey}\inst{2} \and 
        \firstname{Jennifer} \lastname{van Saders}\inst{3}}

\institute{Space Science Institute, 4750 Walnut Street, Suite 205, Boulder CO 80301 USA \and
           Universit\'e C\^ote d'Azur, Observatoire de la C\^ote d'Azur, CNRS, Laboratoire Lagrange, France \and
           Carnegie Observatories, 813 Santa Barbara Street, Pasadena CA 91101 USA}

\abstract{The Kepler mission has been fantastic for asteroseismology of 
solar-type stars, but the targets are typically quite distant. As a 
consequence, the reliability of asteroseismic modeling has been limited by 
the precision of additional constraints from high-resolution spectroscopy 
and parallax measurements. A precise luminosity is particularly useful to 
minimize potential biases due to the intrinsic correlation between stellar 
mass and initial helium abundance. We have applied the latest version of 
the Asteroseismic Modeling Portal (AMP) to the complete Kepler data sets 
for 30 stars with known rotation rates and chromospheric activity levels. 
We compare the stellar properties derived with and without the measured 
parallaxes from the first data release of Gaia. We find that in most cases 
the masses and ages inferred from asteroseismology shift within their 
uncertainties. For a few targets that show larger shifts, the updated 
stellar properties only strengthen previous conclusions about anomalous 
rotation in middle-aged stars.}

\maketitle

%%%%%%%%%%%%%%%%%%%%%%%%%%%%%%%%%%%%%%%%%%%%%%%%%%%%%%%%%%%%%%%%%%%%%%%%%%%
\section{Motivation}\label{sec1}

The correlation between stellar mass and initial helium abundance is a 
long-standing problem in modeling solar-type stars. Increasing either the 
mass or the initial helium yields a model with a higher luminosity, so we 
can trade off one parameter for the other while still satisfying the 
observational constraints \cite{LebretonGoupil2014}. Asteroseismic 
observations can reduce the severity of this problem by providing a strong 
constraint on the stellar radius, but the issue cannot be avoided entirely 
without a more direct constraint on the stellar luminosity. Without such a 
constraint, the correlation can lead to systematic biases in the stellar 
properties inferred from asteroseismology. In this paper, we examine the 
impact of including luminosity constraints derived from Gaia DR1 
parallaxes \cite{Gaia2016} for a sample of 30 Kepler targets with known 
rotation rates \cite{Garcia2014} and chromospheric activity levels 
\cite{Karoff2013}. Our goal is to evaluate the possibility of systematic 
biases in the asteroseismic masses and ages for stars with anomalous 
rotation compared to empirical gyrochronology relations 
\cite{vanSaders2016}, with implications for a new theory of magnetic 
evolution beyond middle age \cite{Metcalfe2016}.

Until recently, it was presumed that rotation and magnetism decay together 
throughout the lives of solar-type stars \cite{Skumanich1972, Barnes2007}. 
Although stars are formed with a range of initial rotation rates, the 
stellar winds entrained in their magnetic fields lead to angular momentum 
loss from magnetic braking. This forces convergence to a single rotation 
rate at a given mass after roughly 500~Myr. The Kepler mission allowed the 
measurement of rotation periods in old field stars whose masses and ages 
could be determined from asteroseismology. These new data revealed a 
population of field stars rotating more quickly than expected from 
gyrochronology \cite{Angus2015}, suggesting that magnetic braking may 
operate with a dramatically reduced efficiency beyond a critical Rossby 
number, Ro\,$\sim$\,2 \cite{vanSaders2016}. A magnetic counterpart to this 
rotational transition was recently identified by \cite{Metcalfe2016}. They 
proposed that a change in the character of differential rotation is the 
underlying mechanism that ultimately disrupts the large-scale organization 
of magnetic fields in solar-type stars. The process begins at 
Ro\,$\sim$\,1, where many global convection simulations exhibit a 
transition from solar-like to anti-solar differential rotation 
\cite{Gastine2014}. This change in the character of differential rotation 
triggers a phase of rapid magnetic evolution leading to a reduction in the 
efficiency of magnetic braking, probably due to a shift in magnetic 
topology \cite{Reville2015}.

% FIGURE 1 ---------------------------------------------------------------- 
\begin{figure*}\label{fig1} 
\centering\includegraphics[angle=270,width=13.5cm,clip]{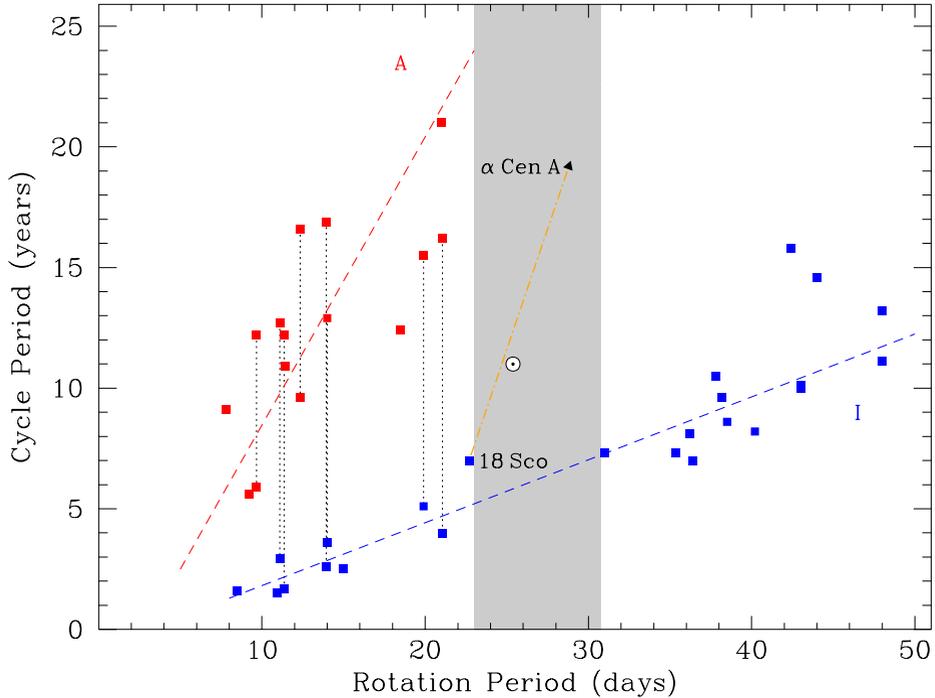} 
\caption{Updated version of a diagram published in \cite{BohmVitense2007} 
using data from \cite{SB1999}, showing the active (A) and inactive (I) 
sequences. Multiple cycles observed in the same star are connected with 
vertical dotted lines. The shaded region indicates the rotation periods 
around the Sun ($\odot$) where Mount Wilson stars do not show cycles.}
\end{figure*}
%-------------------------------------------------------------------------

The new picture of rotational and magnetic evolution provides a framework 
for understanding some observational features of stellar activity cycles 
that have until now been mysterious. An updated version of a diagram 
published in \cite{BohmVitense2007} is shown in Figure~\ref{fig1}, using 
data from \cite{SB1999}. More recent data have been added from 
\cite{Hall2007, DeWarf2010, Metcalfe2010, Metcalfe2013, Ayres2014, 
Egeland2015, Salabert2016}. All of the slower rotators ($P_{\rm 
rot}>30$~days) are K-type stars, which is now understandable---magnetic 
braking ceases in more massive main-sequence stars before they reach these 
long rotation periods. The transition across Ro\,$\sim$\,2 for G-type 
stars occurs at rotation periods comparable to the Sun ($P_{\rm 
rot}\sim$\,23--30~days). If we consider the evolutionary sequence defined 
by 18~Sco, the Sun, and $\alpha$~Cen~A, the data suggest that a normal 
cycle on the inactive sequence may grow longer across this transition 
(dot-dash line) before disappearing entirely.

%%%%%%%%%%%%%%%%%%%%%%%%%%%%%%%%%%%%%%%%%%%%%%%%%%%%%%%%%%%%%%%%%%%%%%%%%%%
\section{Asteroseismology with Kepler}\label{sec2}

The Kepler space telescope yielded unprecedented data for the study of 
solar-like oscillations in other stars. Initial indications of anomalous 
rotation and magnetic activity in old field stars relied on asteroseismic 
properties from \cite{Metcalfe2014}, which were based on an analysis of 
only 9 months of data \cite{Appourchaux2012}. Kepler completed its primary 
mission in 2013, but the large samples of multi-year observations posed an 
enormous data analysis challenge that has only recently been surmounted 
\cite{Davies2015, Davies2016, Lund2017}. The longer data sets improved the 
signal-to-noise ratio (S/N) of the power spectrum for fainter stars with 
previously marginal detections, and yielded additional oscillation 
frequencies for the brighter targets. This expanded the asteroseismic 
sample of stars with known rotation rates \cite{Garcia2014} and 
chromospheric activity levels \cite{Karoff2013}, from 21 targets 
\cite{Ceillier2016} to more than 30.

% FIGURE 2 ---------------------------------------------------------------
\begin{figure*}\label{fig2} 
\centering\includegraphics[angle=270,width=13.5cm,clip]{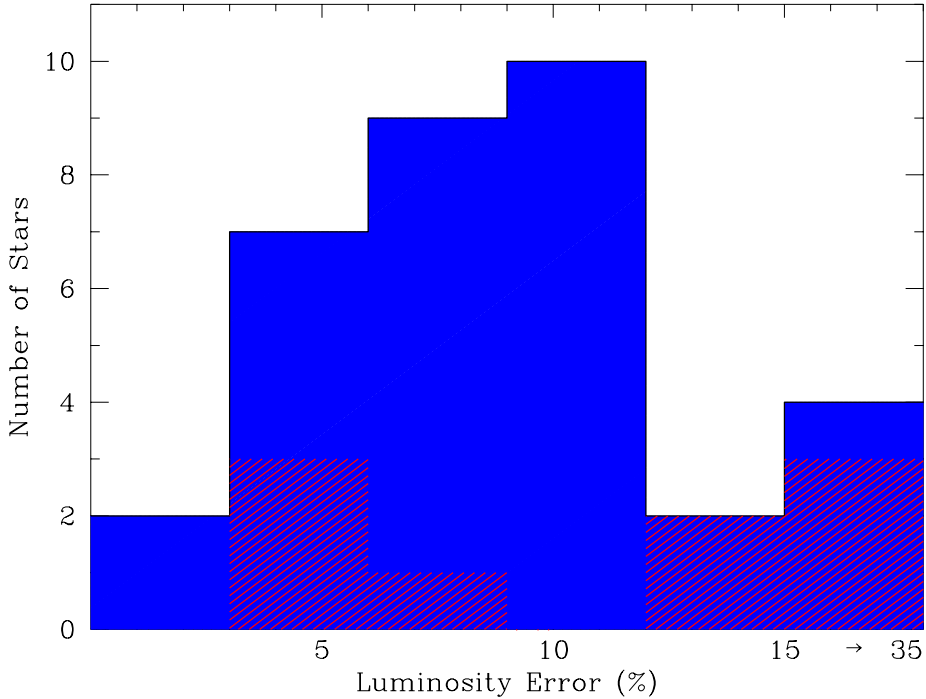} 
\caption{Histogram of the available luminosity constraints for our sample 
from the Tycho catalog (red) and from Gaia DR1 (blue). All of the stars in 
our sample now have luminosity constraints, and most have a precision 
better than 10\%. In late 2017, Gaia DR2 promises to improve the precision 
by nearly an order of magnitude.}
\end{figure*}
%-------------------------------------------------------------------------

We have used the latest version of the Asteroseismic Modeling Portal (AMP) 
\cite{Metcalfe2009} to obtain reliable stellar properties from the 
complete Kepler data sets \cite{Creevey2017}. For each star in the sample, 
we repeated the modeling with and without a luminosity constraint derived 
from Gaia DR1 parallaxes \cite{Gaia2016}, allowing us to isolate the 
impact of this constraint on the inferred stellar properties. Parallaxes 
from the Tycho catalog were previously available for a few of the targets, 
but all of the stars in our sample now have luminosity constraints---and 
most have a precision better than 10\% (see Figure~\ref{fig2}).

%%%%%%%%%%%%%%%%%%%%%%%%%%%%%%%%%%%%%%%%%%%%%%%%%%%%%%%%%%%%%%%%%%%%%%%%%%%
\section{Impact of Gaia parallaxes}\label{sec3}

\begin{figure*}\label{fig3}
\centering\includegraphics[angle=270,width=13.5cm,clip]{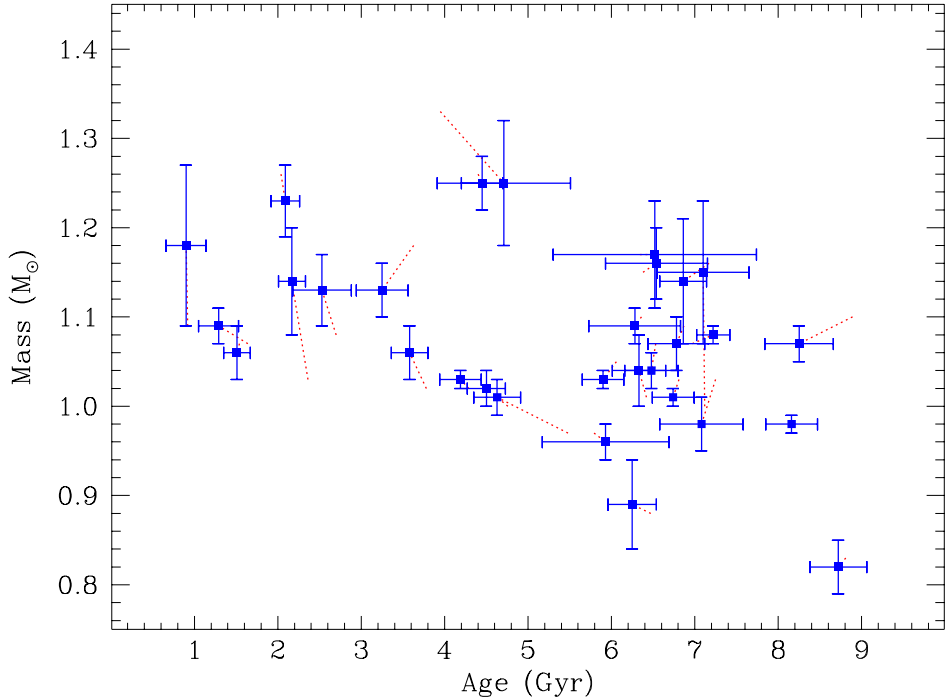}
\caption{Stellar masses and ages inferred from asteroseismology before and 
after including luminosity constraints derived from Gaia DR1 parallaxes. 
Blue points with uncertainties show the values obtained when including a 
luminosity constraint, and dotted red lines connect to the values inferred 
without a luminosity constraint.}
\end{figure*}

Claims of anomalous rotation in old field stars depend most directly on 
the inferred asteroseismic ages. The masses are also important because 
magnetic braking depends on the depth of the surface convection zone, and 
angular momentum evolution ultimately follows changes in the moment of 
inertia as a star expands into a subgiant. In Figure~\ref{fig3} we 
illustrate the impact of the available luminosity constraints on the 
inferred asteroseismic masses and ages. Most of the values shift within 
their respective uncertainties. For a few targets that show larger shifts, 
the updated stellar properties only strengthen previous conclusions about 
anomalous rotation in middle-aged stars.

% FIGURE 3 ---------------------------------------------------------------
%-------------------------------------------------------------------------

In late 2017, Gaia DR2 promises to improve the parallax precision for all 
of our targets by nearly an order of magnitude. This will allow us to 
eliminate any remaining biases in the inferred stellar properties for our 
sample. It may also yield the asteroseismic composition with sufficient 
accuracy to distinguish between competing Galactic enrichment models.

%%%%%%%%%%%%%%%%%%%%%%%%%%%%%%%%%%%%%%%%%%%%%%%%%%%%%%%%%%%%%%%%%%%%%%%%%%%
\begin{acknowledgement}
\noindent\vskip 0.2cm
\noindent {\em Acknowledgments}: This work was supported in part by NASA 
grants NNX15AF13G and NNX16AB97G, and by White Dwarf Research Corporation 
through the Non-profit Adopt a Star program. Computational time at the 
Texas Advanced Computing Center was provided through XSEDE allocation 
TG-AST090107.
\end{acknowledgement}

%%%%%%%%%%%%%%%%%%%%%%%%%%%%%%%%%%%%%%%%%%%%%%%%%%%%%%%%%%%%%%%%%%%%%%%%%%%
\bibliography{metcalfe}
%%%%%%%%%%%%%%%%%%%%%%%%%%%%%%%%%%%%%%%%%%%%%%%%%%%%%%%%%%%%%%%%%%%%%%%%%%%

\end{document}